\documentclass[aps,prl,twocolumn,superscriptaddress,showpacs]{revtex4}
\usepackage{graphicx}
\usepackage{epsfig}
\usepackage{bm}
\usepackage{times}
\usepackage{amssymb}
\usepackage{url,hyperref}

\begin{document}

\title{Coexistence of ferromagnetism and Kondo effect in uranium compounds}

\author{N. B. Perkins}
\affiliation{MPIPKS, N\"othnitzer Str. 38, 01187 Dresden, Germany}
\affiliation{Bogoliubov Laboratory of Theoretical Physics, JINR,
Dubna, Russia}
\author{J. R. Iglesias}
\affiliation{Instituto de F\'{\i}sica, Universidade Federal do Rio
Grande do Sul, 91501-970 Porto Alegre, Brazil}
\author{M. D. N\'{u}\~{n}ez-Regueiro}
\author{B. Coqblin}
\affiliation{Laboratoire de Physique des Solides, Universit\'{e}
Paris-Sud, B\^atiment 510, 91405 Orsay, France}


\begin{abstract}
Coexistence between ferromagnetic order and Kondo behavior has been
observed in some uranium compounds. The underscreened Kondo lattice
model can provide a possible description of this coexistence. Here
we present a model of a lattice of $S=1$ spins coupled to the
conduction electrons through an intra-site exchange interaction
$J_{K}$ and an inter-site ferromagnetic exchange $f-f$ interaction
$J_{H}$. Finite temperature results show that the Kondo temperature
is larger than the Curie ordering temperature, $T_{C}$, providing a
possible scenario for the coexistence of Kondo effect and magnetic
order. Also, the Kondo behavior disappears abruptly for low values
of $J_{K}$ and smoothly when changing the band occupation. These
results are in qualitative agreement with the experimental situation
for the above mentioned uranium compounds.
\end{abstract}


\maketitle

Many rare-earth and actinide compounds show anomalous properties
which have been attributed to heavy fermion behavior. A lot of
attention has been placed on the experimental and theoretical study
of the competition between magnetic order - mainly antiferromagnetic
(AF) - and Kondo effect in cerium or ytterbium
compounds~\cite{CoqblinPM,CoqblinKLM}. As a result, nowadays there
exist various available theories describing many reliable
experiments, or vice versa. For a review see ref.~\cite{CoqblinPM}.

On the other hand, uranium compounds present many distintic
behaviors from mixed valence to magnetic order with or without a
Kondo effect and even a coexistence between ferromagnetic order and
superconductivity, as it has been observed in $UGe_{2}$ and
$URhGe$~\cite{flouquet}.

However, one of the most interesting experimental findings, the
coexistence of Kondo behavior and ferromagnetic order observed in
some uranium compounds has been -- to our knowledge -- somewhat
overlooked from a theoretical point of view. It's a long way since
the first experimental evidence of coexistence between Kondo
behavior and ferromagnetic order in the dense Kondo compound
$UTe$~\cite{Schoenes}. More recently this coexistence has been
observed in $UCu_{0.9}Sb_{2}$ ~\cite{Bukowski} and $UCo_{0.5}Sb_{2}$
~\cite{Tran}. Those systems undergo a ferromagnetic ordering at
relatively high Curie temperatures of $T_{c}$ = 102K, $T_{c}$ = 113K
and $T_{c}$ = 64.5K, correspondingly, and also present a logarithmic
decrease of the magnetic resistivity above it, suggesting a Kondo
behavior at high temperature. These experimental results clearly
show the coexistence of ferromagnetic order and Kondo behavior below
a large Curie temperature. Another example of  such coexistence has
been found in $URu_{2-x}Re_{x}Si_{2}$ compounds, where a non-Fermi
liquid (NFL) behavior of resistivity and specific heat has been
observed inside the ferromagnetic phase~\cite{dav,bauer}.

In order to account for the Kondo-ferromagnetism coexistence
observed in the above quoted uranium compounds, we propose the
framework of the underscreened Kondo lattice (UKL) model, that
incorporates all the essential aspects of the problem such as Kondo
interaction and magnetic RKKY interaction between localized spins
$S=1$. We believe it is reasonable to assume a $5f^2$ configuration
for uranium because the magnetic moments deduced from magnetic
experiments in these compounds are close to the free ion
values~\cite{Schoenes2} and moreover a good agreement with
experiments has been obtained by considering only 2 localized 5f
electrons~\cite{Zwicknagl}.

The underscreened Kondo impurity model has been studied using
different approaches\cite{coleman1} and an exact solution was
obtained by the Bethe-ansatz method~\cite{schlott}. On the other
hand there are relatively few studies of the UKL, among them
refs.[\onlinecite{perkins06,flor}]. Here we describe the UKL as a
periodic lattice of magnetic atoms with $S=1$ -- which can be
modeled by two degenerate $f$-orbitals -- interacting with itinerant
spins. The spins of these $f$-electrons are coupled to $S=1$ due to
the strong on-site Hund's coupling and we treat them on the basis of
a fermionic representation for spin $S=1$ \cite{perkins02}, valid in
this constrained triplet Hilbert space. Then, we perform a mean
field analysis of the model, both at zero- and finite-temperatures
in terms of the Green's functions of the system.

In normal Kondo lattice (KL) model ($S=1/2$, i.e. one $f$-electron
per lattice site) the competition between Kondo effect and magnetic
RKKY interaction leads to the appearance of a quantum phase
transition between a magnetically ordered (generally
antiferromagnetic) phase and a phase with coherent Kondo
spin-singlet formation and short range magnetic
correlations~\cite{Iglesias,CoqblinKLM}. In a similar way, with the
help of the UKL model we can describe quantum ferromagnetic phase
transitions and the coexistence between ferromagnetic order and NFL
behavior~\cite{bauer} at different values of external parameters
such as band filling, pressure or temperature. Based on this
analysis we can draw a phase diagram showing magnetic ordered phases
with no Kondo effect and regions of coexistence between
ferromagnetic order and Kondo behavior.

The UKL Hamiltonian can be written in the following form:

\begin{eqnarray} H =
\sum_{\vec{k},\sigma}(\epsilon_{\vec{k}}-\mu)\mathbf{n}_{\vec{k}\sigma}^{c}+
\sum_{i\sigma\alpha}E_{0}\mathbf{n}_{i\sigma}^{f\alpha}
+\nonumber \\
J_K \sum_{i}{\bf S}_i{\bf{\sigma}_i} + \frac{1}{2}J_H \sum_{ij}{\bf
S}_i{\bf S}_j \label{Hamil}
\end{eqnarray}
where the first term represents the conduction band with dispersion
$\epsilon_{k}$, width $2D$ and a constant density of states $1/2D$,
while $\mu$ is the bare electron chemical potential. Localized spins
are represented by fermionic operators
$\mathbf{f_{i\sigma\alpha}^{\dagger}}$ and
$\mathbf{f_{i\sigma\alpha}}$ introduced in
refs.[\onlinecite{perkins02,perkins06}], carrying spin, $\sigma$,
and orbital, $\alpha$, indexes. $E_{0}$ is a Lagrange multiplier
which is fixed by a constraint for the total number of f-electrons
per site, $n_{f}=\sum_{\sigma}(n_{\sigma}^{f1}+n_{\sigma}^{f2})=2$,
and can be interpreted as a fictitious chemical potential for
$f$-fermions. The third term is the on-site Kondo coupling,
$J_{K}>0$, between localized $\mathbf{S_i}=1$ and  conduction
electron's $\sigma_i=1/2$ spins, whose fermionic representation is
done in the usual way. And finally, the last term is the
ferromagnetic inter-site interaction, $J_{H}<0$ , between localized
$f$-magnetic moments, resulting explicitly from two contributions:
the effective RKKY interaction, and the direct exchange.

As a first step we introduce the relevant fields. They are
$\widehat{\lambda}_{i\sigma}=\sum_{\alpha}
\mathbf{c_{i\sigma}^{+}}\mathbf{f_{i\sigma}^{\alpha}}$ which couple
$c_{i\sigma}$ and $f_{i\sigma}^{\alpha}$ electrons at the same site,
and operators of magnetization for both $c$- and $f$- subsystems,
which are, respectively,
$\mathbf{M_i}=S_i^z=\frac{1}{2}(n_{i\uparrow}^f-n^f_{i\downarrow})$
and
$\mathbf{m_i}=\sigma_i^z=\frac{1}{2}(n^c_{i\uparrow}-n^c_{i\downarrow})$
describing the long range magnetic order of the system. Then we
perform a mean field (MF) analysis of the Hamiltonian eq. (1),
expanding it according to the four following averages:
$\lambda_{\sigma}=\langle\widehat{\lambda}_{i\sigma}\rangle$, $M
=\langle \mathbf{M_i}\rangle$ and $m=\langle \mathbf{m_i}\rangle$.
The non-zero values of $M$ and $m$ correspond to the formation of
the magnetic phase with finite total magnetization, while the
parameter $\lambda_{\sigma}$ describes an effective hybridization
between conduction and $f-$ electrons that corresponds to the Kondo
behavior. In MF approximation the average value of $\sum_{\alpha}
c_{i\sigma}^{+}f_{i\sigma}^{\alpha}$ may be different from zero,
implying spurious charge fluctuations. But one should keep in mind
that the original term of the Hamiltonian is a four-fermion
operator, i.e. a product
$\widehat{\lambda}_{\sigma}\widehat{\lambda}_{\bar{\sigma}}$~\cite{burdin01}.
So, $\lambda_{\sigma}$ should be interpreted as a tool for the MF
calculation that provides a correct description of the Kondo
temperature, for example in the study of the Kondo
impurity~\cite{Lacroix-79}. This procedure is basically equivalent
to other mean field approaches developed for studying normal KL,
such as the path-integral calculation restricted to a saddle-point
solution performed by Coleman and Andrei\cite{coleman}, or
large-{\it N} formulation performed by Burdin et al~\cite{burdin02},
where again the saddle-point solution yields a mean-boson-field
approximation.

The resulting MF Hamiltonian reads
\begin{eqnarray}
\begin{array}{l}
H_{MF}= \sum_{\vec k \sigma} \varepsilon_{\vec k\sigma} n_{\vec
k\sigma}^{c} +
 \sum_{i\alpha\sigma}
E_{0\sigma }n_{i\alpha\sigma }-\\[0.2cm]
-\frac{1}{2}J_K \sum_{i\alpha\sigma} (\lambda_{\bar{\sigma}}
\widehat{\lambda}_{i{\sigma}}^{{\alpha}} + h.c.)
+\\[0.2cm]
2J_K N \sum_{\sigma} \lambda_{\bar{\sigma}} \lambda_{{\sigma}} - J_K
N m  M -\frac{1}{2}J_H N z M ^2
\end{array}
\label{Ham1}
\end{eqnarray}
\noindent where, remembering that $J_K>0$ and $J_H<0$, energies read
\begin{eqnarray}
\begin{array}{l}
\varepsilon_{\vec k\sigma} =\varepsilon_{\vec k}+J_K\sigma M ~,
~with ~~~~\sigma=\pm 1/2
 \\[0.2cm]
E_{0\sigma } =E_0+J_K\sigma m -\frac{1}{2}J_K \lambda_{{\sigma}}
\lambda_{\bar{\sigma}} +J_H z \sigma M
\end{array}
\label{def1}
\end{eqnarray}

The diagonalization of the MF Hamiltonian (\ref{Ham1}) gives two
non-hybridized $f$-bands (one for each spin) at energies $E_{0\sigma
}$ and two quasiparticle bands $E^{\sigma}_{\pm}(\vec{k})$ with
energies
\begin{eqnarray}
E^{\sigma}_{\pm}(\vec{k})=\frac{1}{2}[ E_{0\sigma}+\varepsilon_{\vec
k\sigma}\pm \sqrt{(E_{0\sigma}-\varepsilon_{\vec k\sigma})^2+8
\alpha^2_{\bar{\sigma}}  }] \label{spectrum}
\end{eqnarray}
\noindent where $\alpha_{\bar{\sigma}}=
-\frac{1}{2}J_K\lambda_{\bar{\sigma}}$. The $\pm$ sign refers to the
upper (lower) hybridized band. Let us point out here one of the main
differences between the mean field treatment of the normal KL and
UKL models: in the underscreened case one $f$-localized level
remains non-hybridized and for the other level the resulting $c-f$
effective hybridization is twice that of the normal case. The energy
spectra $E^{\sigma}_{\pm}(k)$ depend on a set of external parameters
such as band filling $n_c$, Kondo coupling $J_K$ and exchange
interaction $J_H$ and a set of internal parameters, $M$, $m$,
$\lambda_{\sigma}$, $\mu$ and $E_0$, which should be calculated in a
self-consistent way. The system of self-consistent equations can be
obtained by keeping constant the number of $f$- and $c$- electrons.
Their expression can be evaluated from the Green functions of the
system by straightforward calculations :
\begin{eqnarray}
\begin{array}{l}
n_{f}^{\sigma}=
\frac{1}{2D}
\int_{-D +\Delta_{\sigma}}^{D +\Delta_{\sigma}} d\varepsilon_{\sigma}
[
n_F(E_{0\sigma})-\\[0.2cm]
n_F(E_{+\sigma})
\frac{\varepsilon_{\sigma}-E_{+\sigma}}
{W_{\sigma}(\varepsilon_{\sigma})}
+
n_F(E_{-\sigma})
\frac{\varepsilon_{\sigma}-E_{-\sigma}}
{W_{\sigma}(\varepsilon_{\sigma})}]\\[0.2cm]
n_{c}^{\sigma}=
\frac{1}{2D}
\int_{-D +\Delta_{\sigma}}^{D +\Delta_{\sigma}}d\varepsilon_{\sigma}
[
-n_F(E_{+\sigma})
\frac{E_{0\sigma}-E_{+\sigma}}
{W_{\sigma}(\varepsilon_{\sigma})}+\\[0.2cm]
n_F(E_{-\sigma}) \frac{E_{0\sigma}-E_{-\sigma}}
{W_{\sigma}(\varepsilon_{\sigma})}]
\end{array}
\label{ncfsigma}
\end{eqnarray}
\noindent  where $n_F(\omega)=\frac{1}{e^{\frac{\omega
-E_F}{T}}+1}$ is the Fermi distribution function,
$\Delta_\sigma=J_K\sigma M $ and
$W_{\sigma}(\varepsilon)=\sqrt{(E_{0\sigma}-\varepsilon)^2+8
\alpha^2_{\bar{\sigma}}} $. From the Green functions we can also
obtain the expression for $\lambda_{\sigma}$ given by:
\begin{eqnarray}
\begin{array}{l}
\lambda_{\sigma}=
\frac{1}{D}\int_{-D +\Delta_{\sigma}}^{D +\Delta_{\sigma}}d\varepsilon_{\sigma}
[n_F(E_{+\sigma})-n_F(E_{-\sigma})]
\frac{\alpha_{\bar\sigma}}{W_{\sigma}(\varepsilon_{\sigma})}
\end{array}
\label{lambdasigma}
\end{eqnarray}

In order to construct a set of self-consistent equations defined by
(\ref{ncfsigma}) and (\ref{lambdasigma}), we substitute the strong
constraint, $n_{if}=2$ in each site, by a softer one for its
average, and set the band filling by the averaged number of
$c$-electrons, $n_c$, which we consider in the usual range of
partial filling, i.e. $0<n_c<1$ :
\begin{eqnarray}
\begin{array}{l}
n_{f} = n_f^{\uparrow}+ n_f^{\downarrow} =2~,~~~  n_{c}
=n_{c}^{\uparrow}+ n_{c}^{\downarrow}
\end{array}
\label{ncf}
\end{eqnarray}

\begin{figure}[t]
\centerline{\includegraphics[width=7.5cm,
clip=true]{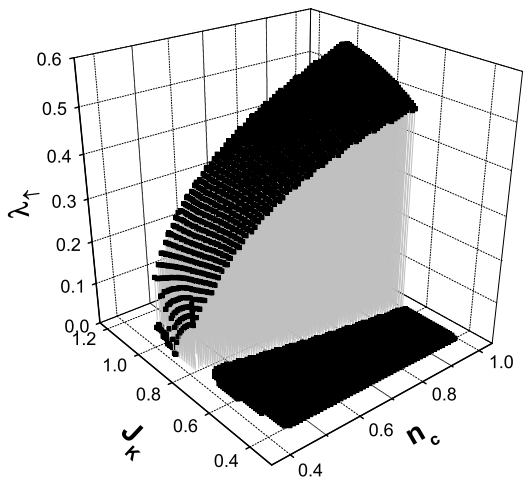}}\caption{Plot of the parameter
$\lambda_{\uparrow}$ as a function of $J_K$ and $n_c$ for $T=0$. The
ferromagnetic interaction between localized $f$-moments is
$J_H=-0.01$. There is a discontinuous transition as a function of
$J_K$ and a behavior in $\sqrt{n_c}$ as explained in the text}
\label{lambda}
\end{figure}

Having solved the system of self-consistent equations and also
minimized the free energy, we study various properties of the UKL
model. In Fig.\ref{lambda} we present a summary of the results for
$T=0$. The region of the parameters $J_K$ and $n_c$ that provide
finite values of $\lambda_\uparrow$ corresponds to a coexistence
between the heavy fermion behavior and ferromagnetic order. The
values of $\lambda_{\downarrow}$ are close to $\lambda_\uparrow$ so
the former is not depicted on the figure, neither the magnetization,
to preserve clearness. It is possible to see on Fig.\ref{lambda}
that $\lambda_\uparrow$ decreases smoothly as a function of $n_c$,
following an approximate square root behavior~\cite{CoqblinKLM},
while it undergoes a sharp transition as a function of $J_K$. When
$\lambda_\uparrow$ goes to zero the ground state is magnetically
ordered with no Kondo effect. The hybridization gap
$\Gamma_{\sigma}=8 \alpha^2_{\bar{\sigma}}$ vanish simultaneously,
signaling a quantum phase transition. Detailed calculations and
results including a full phase diagram with different values of
$J_K$ and $J_H$ will be presented in a forthcoming article.

\begin{figure}[t]
\centerline{\includegraphics[width=7.cm, clip=true]{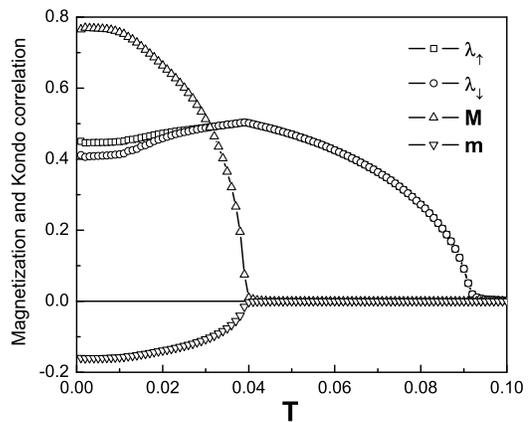}}
\caption{Plot of $\lambda_{\uparrow}$, $\lambda_{\downarrow}$, $M$
and $m$ as a function of temperature. Self-consistent solutions are
obtained for the following parameters:$J_K=0.8$, $J_H=-0.01$ and
$n_c=0.8$ (in units of the half-bandwidth $D$).} \label{temp}
\end{figure}

Finite temperature calculation permits to determine the Curie
temperature and the correlation (Kondo) temperature. Fig.\ref{temp}
shows the behavior of the $f-$ and $c-$ magnetizations, $M$ and $m$
respectively, and $\lambda_{\sigma}$, as a function of the
temperature. The parameters here and in following figures are
$J_K=0.8$, $J_H=-0.01$ and $n_c=0.8$ and all energies and
temperatures are in units of the half-bandwidth $D$. At zero and low
temperatures we observe the coexistence of magnetic order and heavy
fermion behavior, but as both spin systems are strongly polarized
the Kondo effect is concealed. When the magnetization decreases,
$\lambda_{\sigma}$ grows, having its maximum at the Curie
temperature when the magnetization vanishes. Due to the breakdown of
the spin symmetry $\lambda_{\downarrow}$ and $\lambda_{\uparrow}$
are slightly different in the magnetic region but they coincide when
the magnetization vanishes for $T=T_C$, i.e. when the spin symmetry
is restored. For $T>T_C$ the system exhibits only Kondo behavior
($\lambda_\sigma \neq 0$, $M=0$ and $m=0$). Finally, above a
characteristic temperature, the Kondo temperature, $T_K$, both
$\lambda_\sigma$ vanish and the two electron systems are decoupled.

\begin{figure}[t]
\centerline{\includegraphics[width=7.5cm,
clip=true]{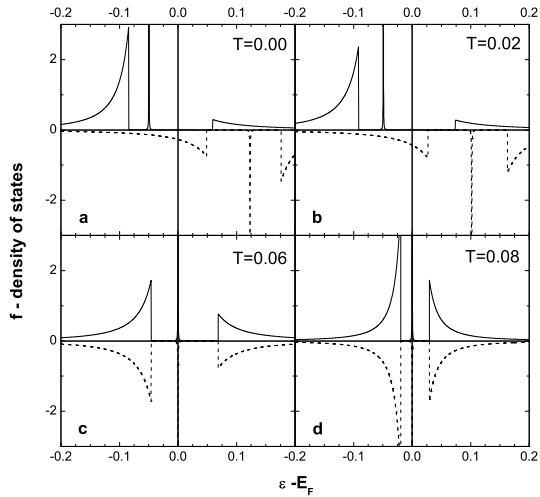}}\caption{ Temperature variation of the up-
(solid line) and down- (dashed line) $f$-density of states, $\rho_{f
\sigma}(\epsilon-E_F)$. Panel (a) corresponds to $T=0$, (b)
$T=0.02$, (c) $T=0.06$ and (d) $T=0.08$. The parameters are the same
as in Fig.\ref{temp}.} \label{density}
\end{figure}

This behavior can be further clarified  by the plot of the densities
of states at different temperatures. As the $c-$densities of states
are almost constant (except in the hybridization gap region) we
present the results only for the $f-$densities of states, which are
calculated numerically from the imaginary part of the $f-f$ Green
functions.In  Fig.[\ref{density}] we plot $\rho_{f
\sigma}(\epsilon-E_F)$ for four different temperatures. At zero and
low temperatures, in the magnetic phase, $T<T_C$, the Fermi level
lies inside the hybridization gap for the up-band, but inside the
conduction band for down-spin, $E^{\downarrow}_{-}(k)$, which
implies a semi-metal behavior in the magnetic phase. Then, an
insulating phase is obtained for $T>T_C$ when the Fermi level is
inside the gap for both up and down spin directions and coincides
with the energy of the f-level, $E_0$, (see, Fig.[\ref{density}c and
d]). Finally, when $\lambda_\sigma$ goes to zero, the hybridization
gap closes and the system becomes metallic, but with a strong heavy
fermion behavior due to the fact that the Fermi level coincides with
the localized f-level, $E_0$.

\begin{figure}[t]
\centerline{\includegraphics[width=8.cm,clip=true]{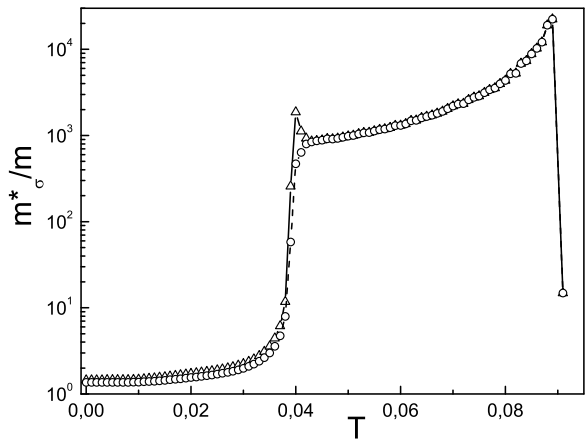}}
\caption{ Temperature dependence of the effective mass enhancement.
The parameter set is  the same as in Fig.\ref{temp}.}
\label{massenh}
\end{figure}

At last, from the quasiparticle spectrum (\ref{spectrum}) one can
estimate the mass enhancement, which will be also
spin-dependent~\cite{bar} :
\begin{eqnarray}
\frac{m_{\sigma}^{*}}{m}= 1 +
\frac{2\alpha^2_{\bar{\sigma}}}{(E_{0\sigma}-E_F-\Delta_{\sigma})^2}
\label{mass}
\end{eqnarray}
We can see in the Fig.\ref{massenh} that for both spin directions
the mass enhancement increases with temperature and becomes dramatic
in the pure Kondo regime, when the magnetization goes to zero. As
the denominator of (\ref{mass}) goes to zero in the pure Kondo
region the effective mass goes strictly to infinite. But in order to
better visualize its behavior and particularly the second transition
at $T_K$, we included a very small finite width in the $f-$level so
obtaining extremely high values for the effective mass in the pure
Kondo region. Hence, two peaks are evident, one corresponds to the
Curie temperature and the second, at higher temperatures, to the
Kondo temperature. In the region of coexistence, the effective mass
increases as a function of temperature following the position of the
Fermi level inside the $E^{\bar{\sigma}}_{-}(k)$ band. Although
simplified, this result provides a qualitatively good explanation of
the behavior of the electronic specific heat.

In conclusion, we have formulated a mean field theory of coexistence
of ferromagnetism and heavy fermion behavior in terms of
self-consistent equations for the relevant fields. Within this
treatment of the UKL model we obtain - for values of $J_K$ between
$0.7$ and  $1.2$ - a region where the order parameters,
$\lambda_\sigma$, the $f-$magnetization, $M$, and the
$c-$magnetization, $m$, are different from zero, characterizing a
coexistence between the heavy fermion (Kondo) properties and
ferromagnetic order. When increasing temperature the magnetic order
disappears at the ordering temperature $T_C$ while $\lambda_\sigma$
is still different from zero, so indicating a Kondo or NFL regime.
This is in agreement with experiments for uranium compounds where a
Kondo behavior is observed also above the Curie temperature, like
the examples we have quoted before. We remark also that the
characteristic Kondo temperature, $T_K$, increases when the Kondo
coupling value $J_K$ also increases. Those results are completely
different from the previous ones obtained for the normal KL model,
where the Kondo effect and the magnetic order are in competition and
cannot be present together ~\cite{Iglesias,CoqblinKLM}. The critical
temperature $T_C$ characterizes here a continuous quantum phase
transition from the ferromagnetic Kondo phase to the non-magnetic
Kondo phase at higher temperatures. Another transition can be
obtained at $T=0$ by varying Kondo coupling parameter $J_K$, but
this time it is discontinuous and leads to a non-Kondo magnetically
ordered state, as it can be observed in Fig.\ref{lambda}. We remark
that this coexistence is mainly due to the presence of both one
non-hybridized and one hybridized $f$-level (see Fig.\ref{density}),
and this ingredient is the key for the existence of the
ferromagnetically ordered Kondo state. Thus, the proposed UKL model
can explain the behavior of some uranium compounds where a
ferromagnetically ordered state coexists with a heavy fermion Kondo
state. The FM-Kondo coexistence arises from the fact that the
localized spins are only partially screened, and consequently they
keep their magnetic interaction at low temperatures.\\
The authors aknowledge stimulating discussions with S. Burdin, P.Fulde,
R. Ramazashvili and J. Schoenes. JRI
thanks MPIPKS, Dresden, and MPIDSO, G\"ottigen, for hospitality. The
authors would like to acknowledge Brazilian agencies CNPq and
FAPERGS for financial support.


\begin{thebibliography}{99}

\bibitem{CoqblinPM} B. Coqblin, M. D. N\'{u}\~{n}ez-Regueiro, Alba
Theumann, J. R. Iglesias and S. G. Magalhaes, Phil. Mag. {\bf 86},
2567 (2006) and references therein

\bibitem{CoqblinKLM} B. Coqblin, C. Lacroix, M. S. Gusmao and J. R.
Iglesias, Phys. Rev. B {\bf 67}, 064417 (2003)

\bibitem{flouquet} J. Flouquet, Progress in Low Temperature Physics,
{\bf 15}, 139-281, ed. by W.P. Halperin, Elsevier (2006)

\bibitem{Schoenes} J. Schoenes, J. Less-Common Met., {\bf 121}, 87 (1986)

\bibitem{Bukowski} Z. Bukowski, R.  Troc, J. Stepien-Damm, C.  Sulkowski and V. H.
Tran, J. Alloys and Compounds, {\bf 403}, 65 (2005).

\bibitem{Tran} V. H. Tran,  R. Troc, Z. Bukowski, D. Badurski and C. Sulkowski,
Phys. Rev. B {\bf 71}, 094428 (2005).

\bibitem{dav} Y. Dalichaouch, M. B. Maple, M. S. Torikachvili and A. L. Giorgi,
Phys. Rev. B {\bf 39}, 2423 (1989)

\bibitem{bauer} E. D. Bauer, V. S. Zapf, P.-C. Ho, N. P. Butch,
E. J. Freeman, C. Sirvent and M. B. Maple, Phys. Rev. Lett. ${\bf
94}$, 046401 (2005)

\bibitem{Schoenes2} J. Schoenes, O. Vogt, J. Lohle, F. Hulliger, and
K. Mattenberger, Phys. Rev. B {\bf 53}, 14987 (1996)

\bibitem{Zwicknagl} G. Zwicknagl, A. Yaresko and P. Fulde, Phys. Rev. B
{\bf 68}, 052508 (2003)


\bibitem{coleman1} J. Gan, P. Coleman and N. Andrei, Phys. Rev. Lett.
{\bf 68}, 3476 (1992).

\bibitem{schlott} P. Schlottmann and P. D. Sacramento, Adv. in Physics
{\bf 42}, 641 (1993)

\bibitem{perkins06} N. B. Perkins, M. D. N\'{u}\~{n}ez-Regueiro,
J. R. Iglesias and B. Coqblin, Physica B ${\bf 378-380}$, 698
(2006).

\bibitem{flor} S. Florens, Phys. Rev. B {\bf 70}, 165112 (2004)

\bibitem{perkins02} S. Di Matteo, N.B. Perkins and C.R Natoli, Phys. Rev. B ${\bf 65}$, 054413
(2002).

\bibitem{Iglesias} J.R. Iglesias, C. Lacroix and B. Coqblin,
Phys. Rev. B ${\bf 56}$, 11820 (1997).

\bibitem{burdin01} S. Burdin,
Thesis, Grenoble (2001)

\bibitem{Lacroix-79} C. Lacroix and M. Cyrot, Phys. Rev. B ${\bf 20}$, 1969
(1979).

\bibitem{coleman} P. Coleman and N. Andrei, J. Phys.: Condens.
Matter ${\bf 1}$, 4057 (1989).

\bibitem{burdin02} S. Burdin, A. Georges and D. R. Grempel, Phys. Rev. Lett. ${\bf
85}$, 1048 (2000)

\bibitem{bar} V. Barzykin, Phys. Rev. B ${\bf 73}$, 094455 (2006).

\end{thebibliography}
\end{document}